\documentclass[manuscript,twocolumn]{aa}
\usepackage{multicol}
\usepackage{graphicx}
\usepackage{txfonts}
\usepackage{epsfig}
\usepackage{times}
\usepackage{natbib}
\usepackage{url}
\usepackage{color}
\usepackage{amssymb,amsmath}
\usepackage{enumerate}
\usepackage[normalem]{ulem}
\usepackage[
  breaklinks = true,
   colorlinks = true,
  urlcolor   = blue,
  citecolor  = blue,
  linkcolor  = red, ]{hyperref}
\graphicspath{ {./figures/} }

\definecolor{manuel}{rgb}{0.0, 0.5, 0.0}
\definecolor{fmi}{rgb}{0.89, 0.0, 0.13}

\usepackage{lineno}

\begin{document}
\title{Interaction of solar jets with filaments: Triggering of large-amplitude filament oscillations}
\author{Reetika Joshi
\inst{1,2},
Manuel Luna
\inst{3,4},
Brigitte Schmieder
\inst{5,6,7},
Fernando Moreno-Insertis
\inst{8,9},
\and
Ramesh Chandra
\inst{10}
}
\institute{Institute of Theoretical Astrophysics, University of Oslo, P.O. Box 1029 Blindern, N-0315 Oslo, Norway
\and
Rosseland Centre for Solar Physics, University of Oslo, P.O. Box 1029 Blindern, N-0315 Oslo, Norway\\
\email{reetika.joshi@astro.uio.no}
\and
Departament F{\'i}sica, Universitat de les Illes Balears, E-07122 Palma de Mallorca, Spain 
\and
Institute of Applied Computing \& Community Code (IAC$^3$), UIB, Spain
\and
LESIA, Observatoire de Paris,  Universit\'e PSL, CNRS, Sorbonne Universit\'e,  Universit\'e de Paris, 5 place Janssen, 92290  Meudon Principal Cedex, France
\and
Centre for mathematical Plasma Astrophysics, Dept. of Mathematics, KU Leuven, 3001 Leuven, Belgium
 \and
SUPA, School of Physics and Astronomy, University of Glasgow, Glasgow G12 8QQ, UK
\and
Instituto de Astrof\'{\i}sica de Canarias, E-38200 La Laguna, Tenerife, Spain 
\and
Departamento de Astrof\'{\i}sica, Universidad de La Laguna, E-38206 La Laguna, Tenerife, Spain 
\and
Department of Physics, DSB Campus, Kumaun University, Nainital -- 263 001, India
}

\authorrunning{Reetika Joshi et al.} 
   \titlerunning{Large amplitude oscillations in a solar filament after jet interaction}
\abstract
{Large-amplitude oscillations (LAOs)  are often detected in filaments. 
Using multiwavelength observations, their origin can be traced back
to the interaction with eruptions and jets.
}
{We present two different case studies as observational evidence in
  support of 2.5D MHD numerical experiments that show that the LAOs
  in the filament channels can be initiated by solar jets. 
}
{ 
We use longitudinal magnetic field observations using the
Helioseismic Magnetic Imager to study the evolution of the filament
channels. The LAOs in the filaments are analysed using two techniques. The
first one is time-distance diagnostics with EUV and H$\alpha$ datasets. 
In the second method, the oscillations in different parts of the filaments are examined using Fourier analysis of the brightness variations of all pixels in H$\alpha$ observations. 
}
%
{
In the two studied events, we can identify a quadrupolar configuration with an X-point at the top of the parasitic region suggestive of a classical null-point. The X-point evolves 
into a flat structure suggestive of a breakout current sheet. A reconnection flow emanates from this structure leading to a jet that propagates along the filament channel.
In both cases we can identify the quiescent and eruptive phases of the jet. 
The triggered LAOs have  periods of around 70-80 minutes and are damped after a few oscillations. The minimum magnetic field intensity inferred with seismology for the filament turns out to be around 30 Gauss.}
{We conclude that the two case studies are consistent with the recent numerical model of \cite{Luna2021}, in which the LAOs are initiated by jets. 
The relationship between the onset of the jet and  filament oscillations is straight-forward  for the first case and less for the second case.
In the second event, although there is some evidence, we cannot rule out other possibilities such as activity unrelated to the null-point or changes in the magnetic structure of the filament.
Both  jets are associated with very weak flares which did not launch any EUV wave. Therefore the role of EUV waves for triggering the filament oscillations can be eliminated for these two cases.}


%

\maketitle 

\section{Introduction}
Solar filaments (or prominences) are clouds of dense and cool plasma hanging in coronal heights \citep{Labrosse2010, Mackay2010, Gibson2018} and their origin is magnetic in nature. Occasionally the active as well as the quiet filaments can oscillate. According to their amplitude, the oscillations can be divided into two categories.
If the amplitude velocity is larger than 10 km s$^{-1}$ they are termed large-amplitude oscillations (LAOs); otherwise they are known as small-amplitude oscillations. 
The oscillations are also grouped according to the relative direction of their motion with respect to the prominence magnetic field.
When the motion is mainly in the assumed direction of the field lines, the oscillations are called `longitudinal'; when perpendicular to them, the oscillation is called  `transverse'.
In this paper we consider only the longitudinal case by studying two large-amplitude longitudinal oscillation (LALO) events.

For a long time now, oscillations in prominences/filaments have been observed.
In early observations they were called winking filaments, as they were moving in and out of relatively narrow bandpasses \citep{Hyder1966,Ramsey1966,Vrsnak1993}. 
However, the first clear detection of a LALO is recent.  It was reported by \citet{Jing2003} where the cold plasma of the prominence oscillated along the filament spine.
Afterward, more cases have been reported \citep[see e.g.,][]{Vrsnak2007,Zhang2012,Luna2012,Luna2014,Luna2017}. Observations show that LALOs have periods of a few tens of minutes up to 160 minutes although they mostly have periods close to one hour \citep{luna_gong_2018}. Recently \citet{luna_extension_2022} have shown theoretically that the period can never exceed 167 minutes. In these oscillations the filament threads move along the dipped magnetic field lines. \citet{Luna2012} proposed a so-called pendulum model where the restoring force is mainly gravity projected along the field lines. The gas pressure gradient also contributes although it is considerably smaller than the gravitational force \citep{Zhang2012,luna_effects_2012}. In the pendulum model the period is given exclusively by the average radius of curvature of the dipped field lines \citep[see also][]{roberts_mhd_2019}. More studies based on 2D and 3D numerical simulations confirmed that the pendulum model describes quite well the longitudinal oscillations \citep{luna_robustness_2016,zhou_three-dimensional_2018,zhang_damping_2019,adrover-gonzalez_3d_2020,liakh_numerical_2020,Liakh2021}.

The trigger for the LAOs is identified variously as Moreton or EUV waves \citep{eto_relation_2002,Okamoto2004, Asai2012, Liu2012, Devi2022}, shock waves \citep{Shen2014}, filament eruptions \citep{Isobe2007, Chen2008}, flares \citep{Vrsnak2007, Li2012}, or jets \citep{Luna2014,Zhang2017,tan_stereoscopic_2023}.
These active phenomena perturb  the filaments, and  produce significant displacements with respect to the equilibrium configuration.
In general, the displacements are larger in LALOs than in the case of transverse oscillations \citep{Schmieder_waves2013,Ofman2015}. \citet{Jing2003} and \citet{Luna2017} reported LALOs with velocity amplitudes on the order of 100 km s$^{-1}$. 

Nowadays it is unclear how the different energetic events can excite the different polarizations of the LAOs. Only a few works have considered this problem.  \citet{liakh_numerical_2020} studied a flux-rope prominence and a very energetic external disturbance. They found that both longitudinal and transverse LAOs are excited. More recently, \citet{Luna2021} studied theoretically for the first time the interaction of a jet with a prominence. They found that jets produce LALOs but also transverse oscillations.
These jets are thought to arise due to the rearrangement of the magnetic field structure in the atmosphere caused either by magnetic flux emergence from beneath the solar surface or due to  driving process of the footpoints of the field lines through photospheric flows, or due to the induced stress in the the coronal magnetic field with a huge release of magnetic energy as a jet or small flare \citep{Joshi2021Balmer, Joshi2021,Joshi2022}. Whatever the  precise magnetic configuration or evolutionary structure leading to the jet eruptions, magnetic reconnection is well accepted to be present at the core of this process for converting the stored magnetic energy to kinetic energy of the jets \citep{Joshi2020, Joshi2020FR, Schmieder_Joshi2022,Schmieder2022}.
   In their model, \citet{Luna2021} explained that the jets are produced in the sheared magnetic arcade near the footpoint of the filament channel. 
   After multiple reconnection processes, the jet flows towards this filament channel and set the filament into oscillations, with amplitude larger than 10 km s$^{-1}$.

    In this paper we present two different observational case-studies of such an interaction between jet and filament, in which the jet initiates the oscillations in the filament channel as suggested by \citet{Luna2021}. 
    The first event on February 5, 2015 concerns an active region (AR) with a huge filament in its periphery.  The AR was the site of repeated flux emergence events and consequently  small flares and jets. For this case, we focus the study on one jet which  hits one end of the filament and initiates long time-period oscillations in the filament structure. The plasma material was pushed along or in the flux rope containing the filament and LAOs were observed. 
    The second event was observed on January 1, 2014 and is chosen from the catalog presented in \citet{luna_gong_2018}. It started with a few small scale brightenings near one footpoint of the filament before the main jet, which flew towards the filament channel and set it into oscillations.  
    In the conclusion we show the similarities between the observations and the model of \citet{Luna2021}.

\section{Description of the Observations and the Oscillation Analysis}
 \begin{figure*}[!ht]
\centering
\includegraphics[width=\textwidth]{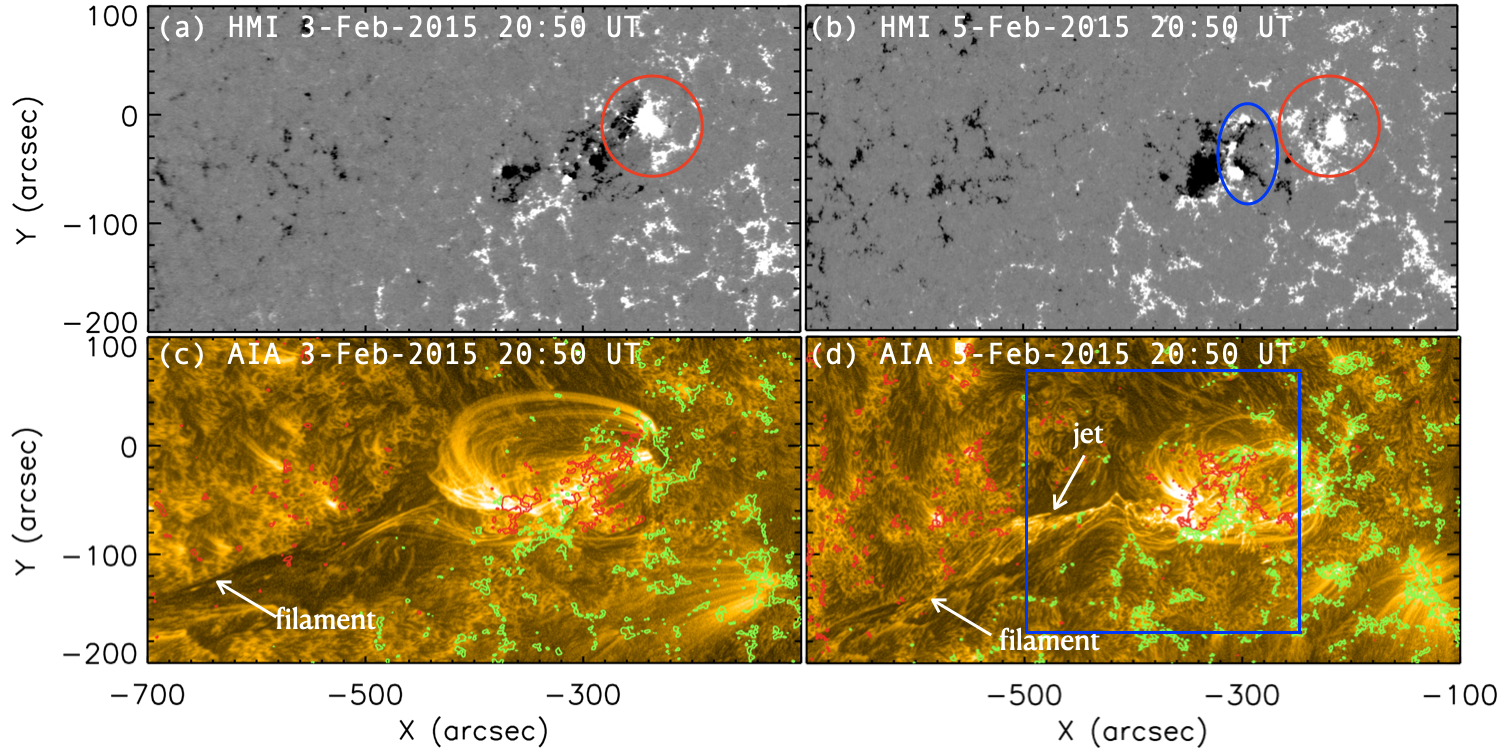}
\caption{Evolution of AR NOAA 12280 on February 3-5, 2015. Top panels: HMI
  magnetograms, bottom panels: AIA 171 images.  The emergence of magnetic
  bipoles is clearly visible in the neighbourhood of the AR and near
  the jet base. Two examples: the red circle in the top panels (a,b)
  indicates a new emerging bipole on February 3; this was followed
  on February 5 by an another emerging bipole marked by the blue
  oval (panel b). The green and red contours are the HMI magnetic field contours of
  strength $\pm$ 100 Gauss respectively. The images are aligned together at
  February 5, 2015 20:50 UT. The blue square is the FOV of Fig. \ref{fig:zoom}.}
\label{fig:general-view-AR-february-5}
\end{figure*}

This study aims to give as complete a picture as possible of the interaction of coronal jets with prominences using data from several ground-based and space observatories. In addition, we also describe the technique to study and parameterise the oscillations in the prominences after jet impact.

\subsection{AIA, H$\alpha$, and HMI observations}
For the study of these two cases, we use the high spatial and temporal resolution EUV data from  the Atmospheric Imaging Assembly \citep[AIA,][]{Lemen2012} instrument on-board the Solar Dynamics Observatory \citep[SDO,][]{Pesnell2012} satellite. AIA  observes the full Sun in seven different UV and EUV wavebands. The pixel size and 
cadence of this instrument are $0.6\arcsec$ and $12$~s respectively. 
In order to increase the contrast of the images, we have deconvolved the EUV data using the AIA point-spread function with the routine  \emph{aia$\_$deconvolve$\_$richardsonlucy.pro} available in solarsoft \citep{Freeland1998}.
All the AIA images are coaligned after  correcting for the solar rotation and are plotted in  logarithmic intensity scale. To  get the best contrast for the filament and jet evolution, the multi-scale Gaussian normalization processing technique \citep[MGN:][]{Morgan2014} is applied to the AIA images. 

For the fulldisk observations of the Sun from ground-based telescopes, we use the  H$\alpha$ line-center photospheric data from the National Solar Observatory/Global Oscillation Network Group (NSO/GONG). To cover the Sun for the entire time of the day, there are six network telescopes situated all over the globe (\url{http://gong2.nso.edu}).
The pixel size and temporal resolution of the H$\alpha$ data are $1$\arcsec
and $1$~min, respectively. 
The filaments and their motions are most easily observed and tracked in H$\alpha$.
We use GONG's H$\alpha$ images to support the AIA EUV observations. H$\alpha$ images show the dark filaments seen in absorption in sharp contrast to the bright chromosphere around them.

 In this study we use the longitudinal magnetic field to show the evolution of the filament channels observed with the Helioseismic Magnetic Imager \citep[HMI,][]{Schou2012}  on-board  SDO.
HMI provides the photospheric magnetograms with  pixel size of $0.5\arcsec$
and a cadence of 45~s 
For both of the studied events, we trace the magnetic history of the associated AR from previous few hours till the start time of the jet filament interaction. Using the HMI magnetic field contours over-plotted on the EUV images, we identify the magnetic reconnection location near the jet base.  

\subsection{Oscillation analysis technique and prominence seismology}\label{subsec:oscillation-analysis}
In the two events described in this work we have used the time-distance diagram technique as in \citet{Luna2014,Luna2017}.
In the first event, we used the GONG H$\alpha$ and the  171 \AA\ SDO/AIA data (Sect. \ref{sec:oscFeb}), whereas, in the second event, we used 171 \AA\ exclusively (Sect. \ref{sec:oscJan}).

We have selected the 171\AA\ passband among all the AIA instrument  filters because this passband shows  not only the absorption  due to the presence of the cool plasma prominence but also the prominence emission thought to be part of the prominence-corona transition region (PCTR) at temperatures $>4 \times 10^5$ K \citep{parenti_nature_2012}. The PCTR is highly filamented, and thus the 171 \AA\ band, with its combination of prominence absorption and emission, can highlight fine structure allowing the tracking of the oscillations: when laying an artificial ``slit'' that crosses the moving filament (as detailed in Sect~\ref{sec:oscFeb} and \ref{sec:oscJan})
and plotting the brightness along the artificial slit by means of a time-distance diagram, clear oscillatory patterns become apparent. Hereafter we will use the term slit instead of an artificial slit. Note that when we use the term ``slit'' we refer to the path we use to measure the intensity and not to the slit of a spectrograph. The oscillatory curves can then be fitted to a damped sinusoidal curve of the form

\begin{equation}\label{eq:sinusoid}
    s(t)= s_0 + A \, e^{-t/\tau} \sin \left( \frac{2 \, \pi}{P} t + \varphi_0 \right) \, ,
\end{equation}
where $s$, $A$, $P$, $\tau$, and $\varphi_0$ are the coordinate along the slit, amplitude, oscillation period, damping time, and  the initial phase respectively.

The study of oscillations with time-distance diagrams is restricted exclusively to a small portion of the filament along which we have traced the oscillations.
However, as we will see, the jet reaches a large part of the filament and makes it oscillate. So, to understand how the different parts of the filament oscillate beyond what is captured by the slit, we will carry out a Fourier analysis of the brightness variations of all pixels in the GONG H$\alpha$ images, using the technique recently developed by \citet{luna_automatic_2022}.  

With this method we will see which regions oscillate and with which period. However, with this technique we cannot extract the rest of the oscillation parameters such as amplitude, damping time or direction of motion.

\cite{Luna2012} showed that the period of the longitudinal oscillations is determined by the curvature of the dips that hold the plasma of the prominence  in the so-called pendulum model \citep[see also][]{Zhang2012}. The reason is that the dominant restoring force in the oscillation is gravity projected along the field lines \citep{luna_effects_2012}. Recently, \citet{luna_extension_2022} have extended the pendulum model by considering that the gravity is not uniform but changes its direction with the position over the solar surface. This introduces important effects for periods larger than 60 minutes. In addition, the authors have found that the periods of the longitudinal oscillations cannot be longer than $P_\odot =$ 167 minutes.

They explained the relation between the period $P$ and the curvature of the dips $R$, as
\begin{equation}\label{eq:new-pendulum}
    \frac{R}{R_\odot}=\frac{\left(P/P_\odot\right)^2}{1-\left(P/P_\odot\right)^2} \, ,
\end{equation}
where $R_\odot$ is the solar radius.
The authors also determined a minimum value of the magnetic field of the dipped lines in order to have magnetic support of the cool mass. Using Eq. (27) from \citet{luna_extension_2022} we obtain
\begin{equation}\label{eq:seismology-magnetic-field}
    B (\mathrm{Gauss}) = \left(47\pm 24\right) \frac{P/P_\odot}{\sqrt{1-\left(P/P_\odot\right)^2}} \, .
\end{equation}
The term in the parenthesis is computed assuming a density range typical for prominences of $\rho=2 \times 10^{-11} - 2 \times 10^{-10} \mathrm{~ kg ~ m^{-3}}$ \citep{Parenti2014}. We consider the prominence density to be the most important source of uncertainty in the estimated magnetic field, because the uncertainties associated with the fit are smaller than the above density range. In the present work, we use these equations to compute the radius of curvature and magnetic field.

\section{Event 1: February 5, 2015}\label{sec:february-5}

The jet-filament interaction on February 5, 2015 started with the ejection of a solar jet parallel to the long filament near the AR 12280. This jet was associated with a circular flare in the AR and it set the filament into oscillations (see the attached animation \url{AIA171_feb.mov}).

\begin{figure}[!h]
\centering\includegraphics[width=0.45\textwidth]{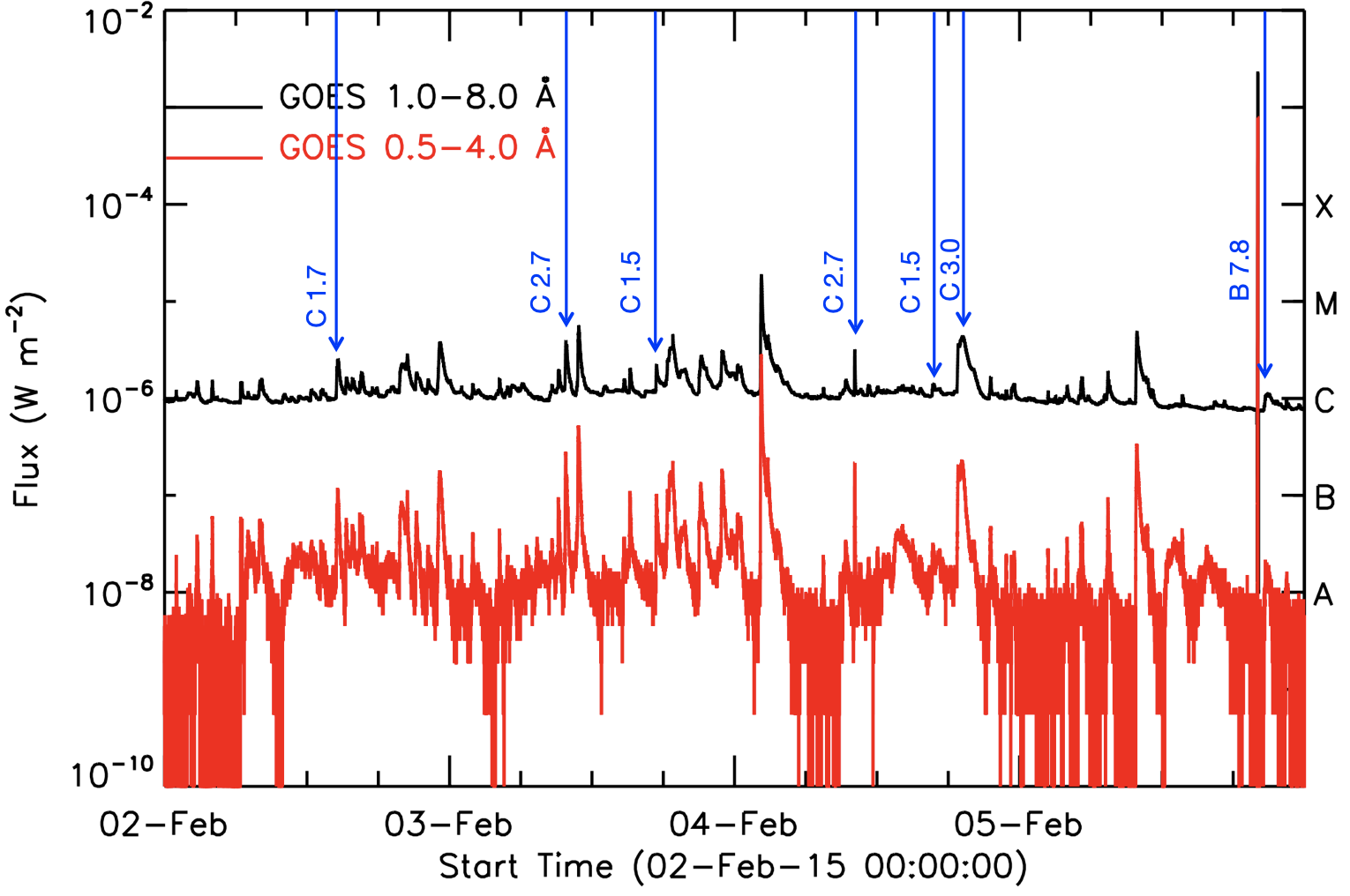}
\caption{Succession of flares in AR 12280 from February 2-5, 2015 registered by GOES. The jet eruption presented here is associated with the GOES B7.8 class flare on February 5 at 20:52 UT.}
\label{Febgoes}
\end{figure}

\begin{figure}[!h]
\centering
\includegraphics
[width=0.5\textwidth]{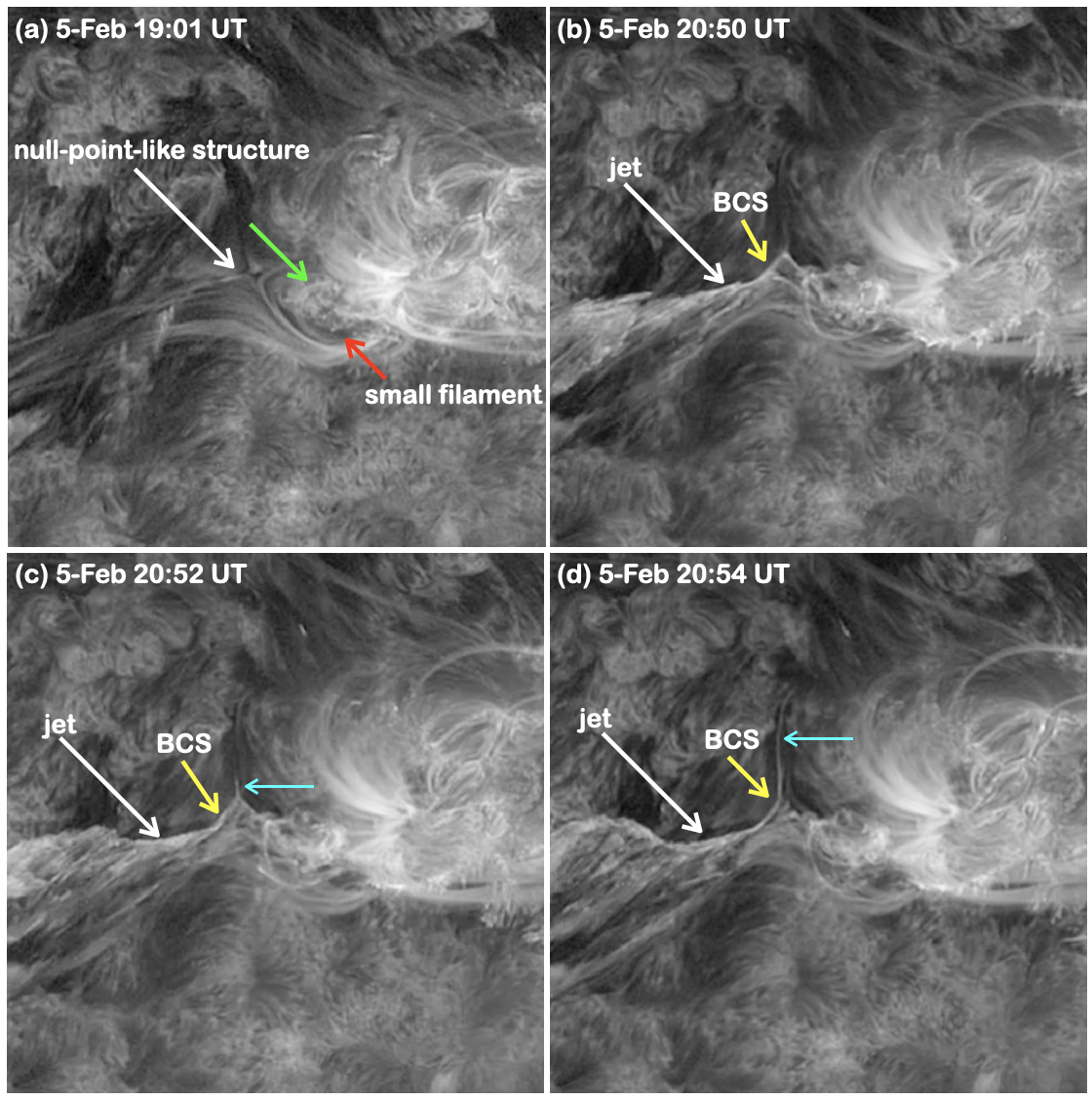}
\caption{Zoom image of the jet base in AIA 171 \AA. The FOV for this figure is presented as the blue square in Fig. \ref{fig:general-view-AR-february-5} (d). The null-point-like structure at the jet base is indicated with a white arrow
(panel a). The jet is initiated under the arcade (green arrow) following a
small filament eruption (red arrow). This small filament pushes the arcade
toward the null-point like structure which leads to the jet flows towards the
filament (white arrow in panel b-d) in the South-West. This flow is bidirectional, as we
observe the second branch, which flows toward the North, is indicated with
cyan arrows in panels (c-d). The null-point type structure collapses into the breakout current sheet (BCS). The latter is marked using yellow arrows in panels b-c-d. (See also the associated animation~\url{AIA171_feb.mov}).  
}
\label{fig:zoom}
\end{figure} 

\begin{figure*}[!h]
\centering
\includegraphics
[width=\textwidth]{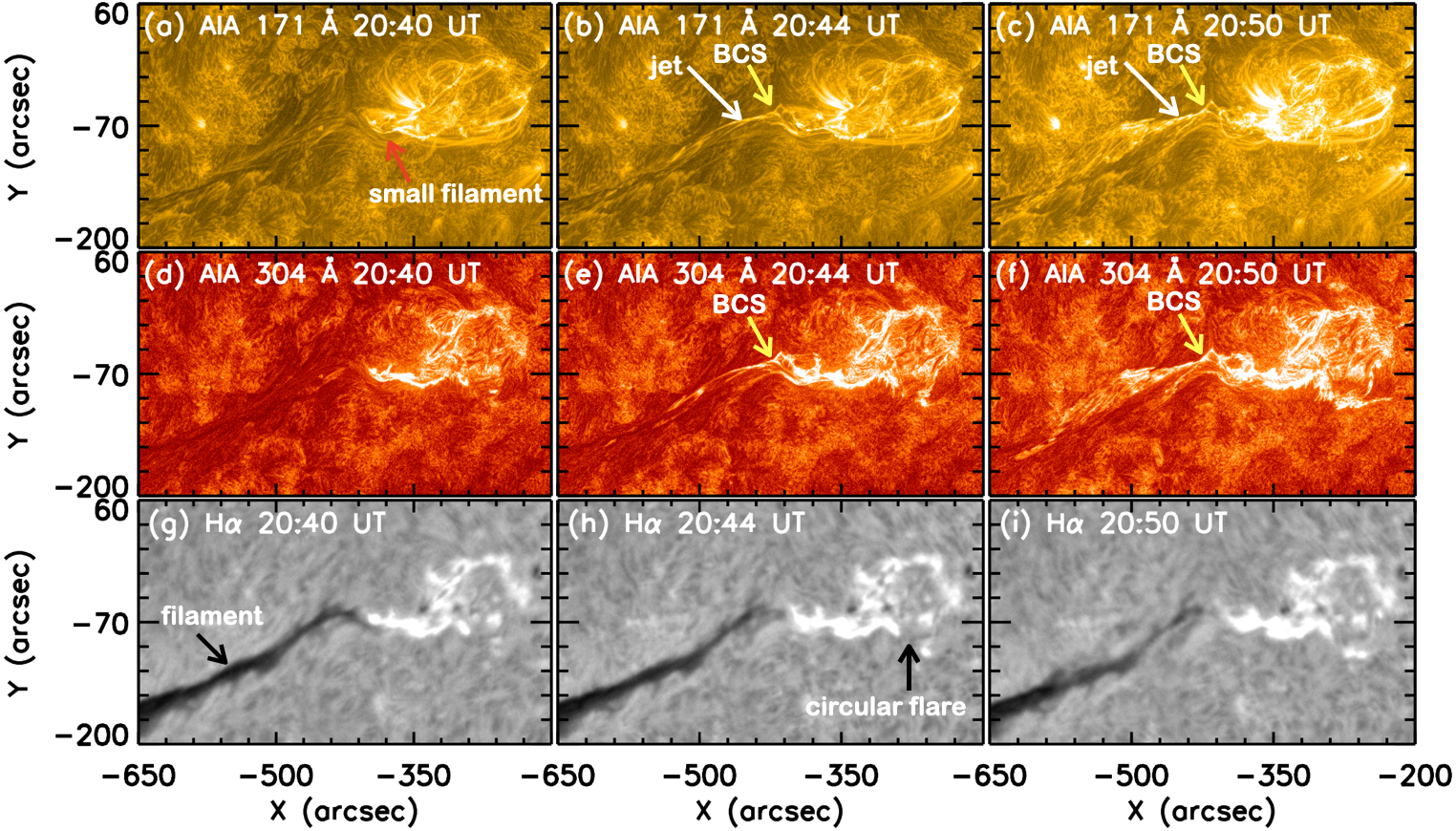}
\caption{
Filament and jet evolution for event 1  on February 5, 2015 observed  in AIA 171 \AA\ (top row), in AIA 304 \AA\ (middle row), in H$\alpha$ (bottom row). 
The jet ejects from a dome-like structure with a small filament eruption shown in panel a.
The jet flows towards the long filament through the BCS 
shown in panel b, c, e, and f and a zoomed view is presented in Fig. \ref{fig:zoom}. The long filament is well observed in all these channels lying as a cool and dense structure (pointed in panel (g)). This jet is associated with a GOES B7.8 circular flare shown in panel h.}
\label{Feb2}
\end{figure*}

\subsection{Environment of the filament and jet}
This  event concerns  active region 
NOAA AR 12280, which appeared on the solar disk on February 2, 2015,  and was located at S08 E17 (-282$\arcsec$, -33$\arcsec$) on February 5   (Fig. \ref{fig:general-view-AR-february-5}). A succession of C-class X-ray flares, related to several magnetic flux  emergence events  was  registered by GOES during its disk passage (see Fig.~\ref{Febgoes}): on February 2 (C 1.7 class flare), on February 3 (C 2.7 and C 1.5 class flares), on February 4 (C 2.7, C 1.5, C 3.0 flares)  and  a B7.8 class flare starting at 20:44 UT and peaked at 20:52 UT on February 5 (\url{https://www.spaceweatherlive.com/en/archive/2015/02/05/xray.html}).
 These flares were accompanied by recurrent jets. We concentrate our study on the last event on February 5 at 20:44 UT.

The magnetic configuration  of the AR consists of a complex polarity pattern with a leading positive polarity
(top panels of Fig. ~\ref{fig:general-view-AR-february-5}).
Successive new episodes of magnetic flux emergence were  observed inside a  remnant bipolar  AR, consequently  increasing 
the complexity of AR 12280. We show two examples: on February 3 there is a clear emerging bipole (red circle  in panel a), and on February 5 a second emerging flux event is indicated by a blue oval in panel b. A large intermediate filament was located at the South-East of the AR along an extended polarity inversion line (PIL). The bottom row of Fig. \ref{fig:general-view-AR-february-5} shows the AR in 171 \AA\  for February 3 and February 5, 2015, where the filament is seen as a dark structure (shown with the white arrow). The filament has one end located over the AR near the origin of the jet. The jet initially pushes a large part of the filament 
and we see periodic motions of the mass which moves back and forth along the filament spine (see movie \url{AIA171_feb.mov}).

\subsection{Generation and Evolution of the
  jet}\label{subsec:evolution-jet20140205} 
The jet was associated with a circular flare and a bidirectional flow towards
the East and North directions. Fig.~\ref{fig:zoom} shows a zoom view
around the jet. The figure contains deconvolved AIA 171 \AA\ images; we use a
black and white representation to increase the contrast. In
Fig~\ref{fig:zoom}a a dome- or arcade-shaped structure overlying the
emerging bipole can be clearly identified with an `X'-point
  configuration at its top,
strongly suggestive of a null-point structure 
(white arrow). A small filament (indicated by a red arrow in 
Fig. \ref{fig:zoom}a) located under the magnetic arcade starts to erupt at
20:30 UT. During its eruption, it encounters the null-point type configuration and produces its collapse 
into a flat structure suggestive of a breakout current sheet (BCS) (marked with yellow arrows in
  panels b through d).  The configuration
  in panel b (20:50~UT) roughly corresponds to the time when the jet flows
reach the filament; the succession of events leading to this configuration 
can be studied in more detail by comparing, as done in 
  Figure~\ref{Feb2}, the evolution in three different filters (AIA 171 \AA,
  AIA 304 \AA and H$\alpha$, top, middle and bottom rows) at times 20:40~UT,
  20:44~UT and 20:50~UT, in the left, center and right columns. The main
  reconnection flow issuing from the BCS leads to a large jet 
that impinges onto the filament. This jet is apparent
as a bright spire issuing from the BCS 
in the AIA 171 and AIA 304 filters at 20:44~UT and 20:50~UT
(Figure~\ref{Feb2}b, c, e and f). 
In the H$\alpha$ images (bottom row) the filament is seen to
  roughly coincide in position with the hot jet; also, the associated circular
  flare in the vicinity of the jet region is clearly visible in those three
  panels. Going back to Fig.~\ref{fig:zoom}, we can see some
  additional consequences of the jet ejection at times a little later than
  those shown in Fig.~\ref{Feb2}: in particular, we can see that, apart from
  the main reconnection flow leading to the large jet, there is a secondary
  flow issuing from the BCS and oriented in the North direction, as marked
  with cyan arrows in panels (b) through (d) of the figure.

\begin{figure}[!h]
\centering\includegraphics[width=0.4\textwidth]{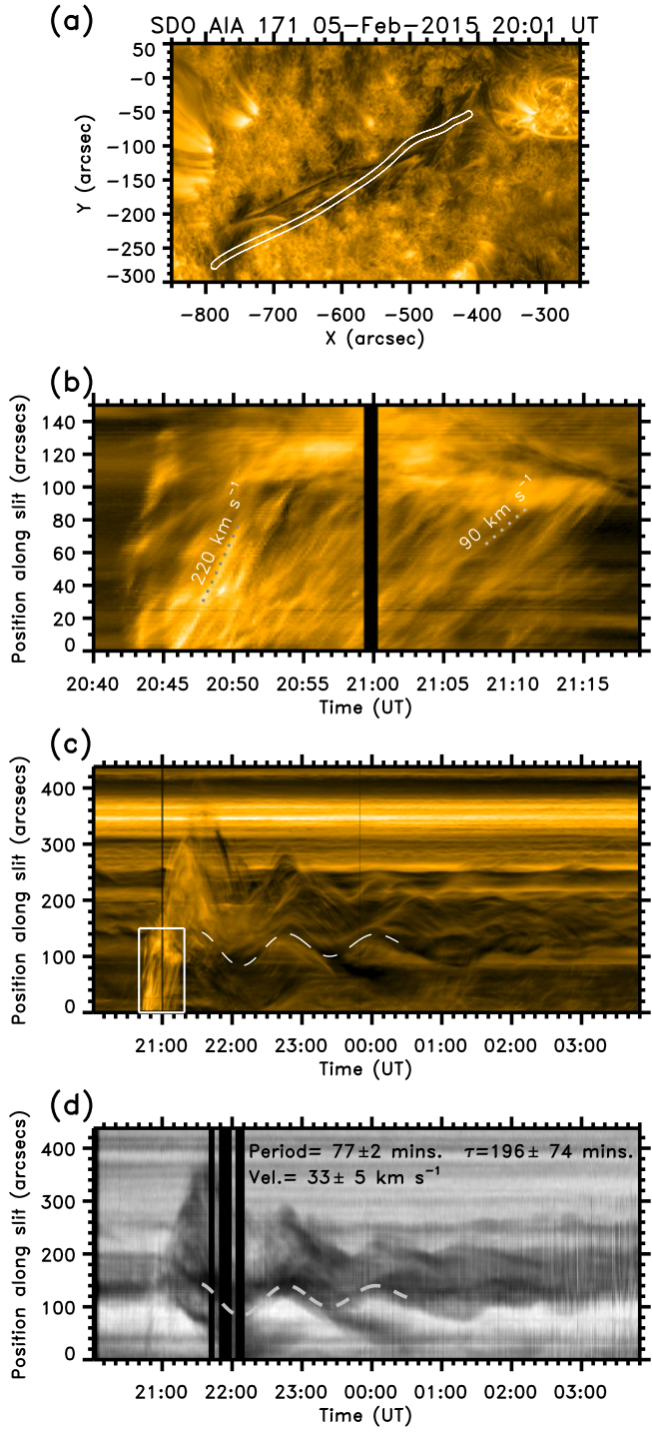}
\caption{Panel (a): Filament seen in absorption in AIA 171 \AA\ on
    February 5, 2015, with the artificial slit used to construct the
    time-distance diagrams (panels b-d) shown as a white
    contour. (Panel b) time-distance diagram for the AIA171 image along the
    topmost $150$~\arcsec of the slit; panel (b) in Fig.~\ref{fig:zoom},
    taken at time 20:50 UT.
    Panels c and d contain the time-distance diagrams for the brightness in  along the curved slit, clearly showing the oscillatory patterns analyzed
    in the paper. The white tracks in panels c and d show one of the fits
    obtained for the oscillation which has
period~$77$~min, velocity amplitude~$33$~$\mathrm{km \, s^{-1}}$ and 
damping time~$\tau=196$ min. The black vertical stripes in panels (b), (c) and (d) are data gaps.
\label{fig:timedistances-event20150205}
}
\end{figure}

The main jet moving towards the filament 
sets  it into oscillations.  
To study the propagation of the main jet and the filament oscillations, 
time-distance diagrams are used 
(Fig~\ref{fig:timedistances-event20150205}). The slit used for the diagrams is shown in panel (a) 
on the background of an AIA 171 \AA image; The choice of this non-straight slit is such that 
  its orientation follows the jet flow and direction of the plasma oscillations of the prominence.
  To match the general direction of the flow, distances along the slit are measured starting at the end nearest to the AR. 
     Leaving the analysis of the
    oscillations to Sec~\ref{sec:oscFeb}, we here focus
    (Fig.~\ref{fig:timedistances-event20150205}b) on the flows apparent
    in the time distance diagram in the region of the slit nearest to the
    BCS, or, more precisely in the $150$\arcsec\ starting at the top-right of
    the slit.
In the figure, we can see a complex structure with lots of bright paths. The slope of the
bright paths is proportional to the flow velocity. 
The dashed lines indicate the trajectory of two such paths, one toward the beginning and the other toward the end of the jet ejection; the first one starts at 
20:48 UT and has a projected speed of about $220$~km~s$^{-1}$ while the second starts
at 
21:08 UT and has a projected speed of about $90$~km~s$^{-1}$.  From those values, and
from direct inspection of the inclination of the tracks in the figure, we
gather that the jet, which always follows  the same trajectory, decreases its velocity (and most probably also its
kinetic energy flux) as time advances.

\subsection{Filament oscillations}
\label{sec:oscFeb}
 To study the oscillations caused by the impact of the jet, the
  time-distance diagram introduced just above has to be extended to the full
  length of the slit and to a much longer time interval than shown in
  Fig~\ref{fig:timedistances-event20150205}(b). The diagram used is shown in
  panel~\ref{fig:timedistances-event20150205}(c), with the white box
  marking the limited domain used in
  panel~\ref{fig:timedistances-event20150205}(b).

The jet flows reach the filament at 20:50 UT as we have already seen in
Sect.~\ref{subsec:evolution-jet20140205}.  The interaction 
produces a displacement of the prominence plasma along the slit. During the
first $50$ minutes we see bright and dark features moving along the slit. We
could say that this is the triggering phase where the jet flow interacts with
the filament plasma. This plasma moves until it reaches maximum elongation at
21:20 UT approximately. Between this time and 02:00 UT 
a clear oscillatory pattern can be identified in panel
(c). However, the movement of the 
prominence plasma is very complex with several specially prominent
oscillation paths visible: the plasma probably oscillates at various levels with different periods and directions.
Due to the difficulty of tracking the movement of a single path we also use
the H$\alpha$ data. Figure \ref{fig:timedistances-event20150205}(d) shows the
H$\alpha$ diagram constructed with the same slit. We can see that 
H$\alpha$ shows a bulk motion of the plasma that is consistent with an
oscillation. We define a curve in the centre of the dark band and then fit
the oscillation using Eq. \eqref{eq:sinusoid}. The resulting fit is plotted
in both panels (c) and (d). The fit is very good in the H$\alpha$ diagram. In
contrast, in AIA 171 \AA\ the curve does not match so well any of the paths
in the diagram. The adjusted period is 77$\pm$2 minutes, however in AIA 171 \AA\ 
it seems that the period of the different threads ranges up to 120 minutes. 

The variation with height of the oscillation period is known from a theoretical point of view \citep{luna_robustness_2016}. 
A similar behaviour was found by \citet{raes_observations_2017}. They pointed out that this effect could produce the false impression of vertically propagating waves. They called these apparent waves \textit{superslow modes.}

Beyond February 6, 02:00 UT the oscillation disappears, as the filament configuration changes and the oscillation moves out of the slit.
The rest of the oscillation parameters from the fit are
the velocity amplitude of $33\pm5$~$\mathrm{km \, s^{-1}}$ and the damping
time~$\tau=196\pm74$~minutes. The ratio between the damping time and the
oscillation period is $\tau/P=2.5$, which indicates that the oscillation
damping is relatively strong. \citet{luna_gong_2018} found typical ratios
$\tau/P=1.25$ for LAOs indicating strong damping of this kind of
motions. Using Eq.~\eqref{eq:new-pendulum}, for a period of 77 minutes we
obtain a radius of the dips $R=188$~Mm. In addition, with 
Eq.~\eqref{eq:seismology-magnetic-field} we can also obtain a lower limit for the magnetic field intensity in the range $B=12-37$ Gauss.  

As, we have just seen, the oscillations in the filament are quite complex
with several periodicities involved. Moreover, the analysis of the
oscillations using a single slit is limited to a small area of the
filament. To 
make a more complete study of what the periodicities are and which regions of
the filament oscillate after the impact of the jet, we applied a technique
developed by \citet{luna_automatic_2022}, as already
mentioned in  Sect. \ref{subsec:oscillation-analysis}.
The technique studies the periodic fluctuations of every pixel of the H$\alpha$ images from the GONG network. With the fast-Fourier transform the power spectral density (PSD) is computed. To discriminate spurious fluctuations from the real oscillations, it is required that the peak in the PSD be greater than several times the background noise so that the confidence level is above 95\%.

\begin{figure}[!h]
\centering\includegraphics
[width=0.45\textwidth]{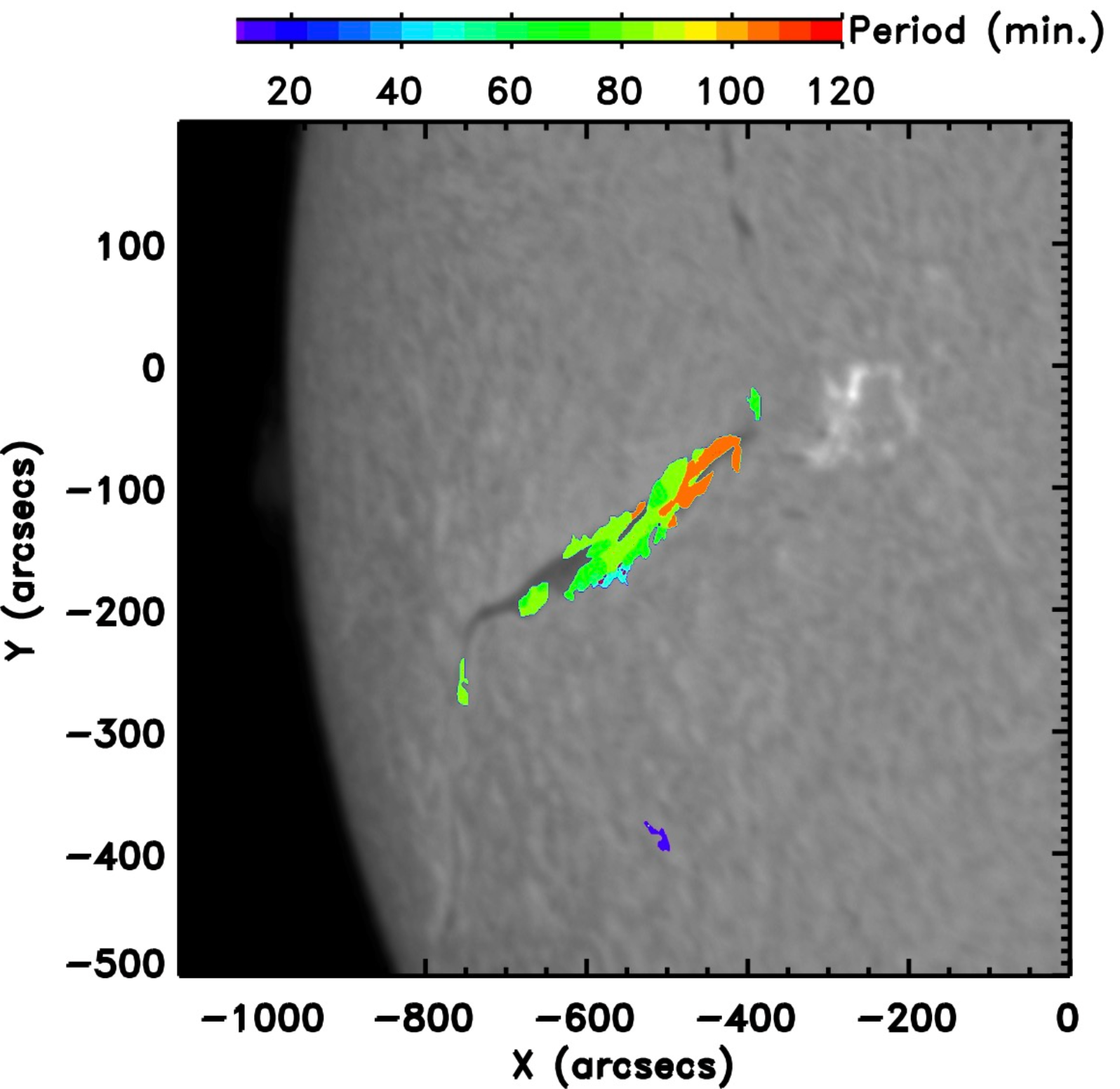}
\caption{Time periods of the filament oscillations over a GONG H$\alpha$ image on February 5, 2015 using the technique developed by \citet{luna_automatic_2022}. 
The color code shows the periods in minutes. \label{fig:periods-over-filament}}
\end{figure}
As shown in Fig. \ref{fig:periods-over-filament}, the jet produces
oscillations in a large part of the filament. The periods range from
$60$--$110$~minutes, which is in agreement with the $77$~minutes obtained with
time-distance method. It is interesting to note that at the northern end a
large area oscillates with periods around $110$~minutes. The
spatial distribution of periods may be associated with the fact that we
observe the oscillations along different field lines and these have different
curvatures in their dips. Using Eq.~\eqref{eq:new-pendulum}, from
  the period range from 77-120 minutes, we obtain that the radius of the dip is in the
  range $103$--$534$~Mm. Similarly, using
Eq.~\eqref{eq:seismology-magnetic-field} we obtain that the field
strength in the filament ranges between $9$ and $64$~Gauss.

\subsection{Comparison with the theoretical model of \cite{Luna2021}}\label{subsec:comparison-model-1st-event}

\begin{figure}
\centering
\includegraphics[width=0.5\textwidth]{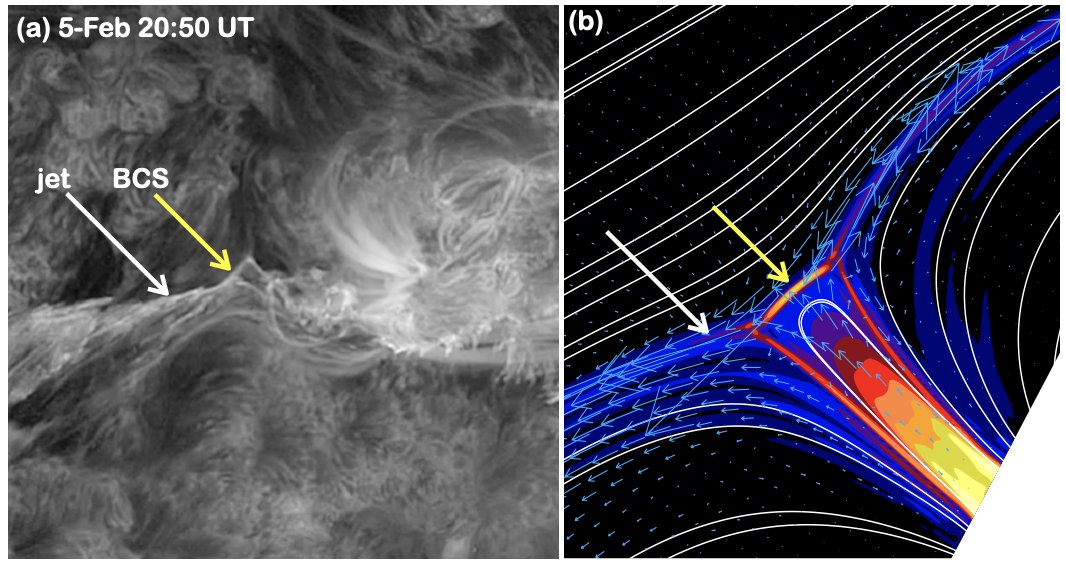}
\caption{Similarity between the observation of event 1 on February 5, 2015 and
  the numerical model of \citet{Luna2021}. The left panel is identical to
  panel b of Fig.~\ref{fig:zoom}; the right panel shows the electric current
  distribution of the model of \citet{Luna2021} (adapted from their
  Fig. 4c). The flat breakout current sheet is marked in both panels by a
  yellow arrow.
  \label{fig:comparison_theoretical_model_electr_current}}
\end{figure}

The observations presented in Sect.~\ref{sec:february-5} bear remarkable
similarities to what was described in the theoretical model of
\cite{Luna2021}. In their model, the magnetic arcade associated with a
bipolar parasitic region at the flanks of the filament channel (FC) of a
prominence was sheared at its photospheric feet. The null point at the top of
the parasitic region initially has a classical X-point shape; through the
photospheric shear, however, the X-point evolves into a breakout current
sheet (BCS) where reconnection takes place; the change in the magnetic
configuration and associated reconnection lead to a succession of nonlinear
waves or jets (Alfvénic front, sonic front, eruptive jet) which propagate
along the filament channel eventually impacting the prominence. An important part of the basic features of that
model seem to be discernable in the observations. Starting with the null
point and BCS, the time series of events in AIA 171 \AA (Fig.~\ref{fig:zoom} and
associated movie \url{AIA171_feb.mov}) seem to show precisely this kind of
configuration: in Fig.~\ref{fig:zoom}a the AIA 171 \AA image clearly delineates
an X-point which then (panels b and c) evolves into a flat BCS. In
Fig.~\ref{fig:comparison_theoretical_model_electr_current}, we compare the
electric current distribution at a relevant time in the numerical model, with
its characteristic flat top, to Fig.~\ref{fig:zoom}b. There is a clear
morphological and evolutionary similarity between the AIA 171 \AA brightness
patterns and the results of the model. 

\begin{figure}[h!]
\centering
\includegraphics[width=0.5\textwidth]{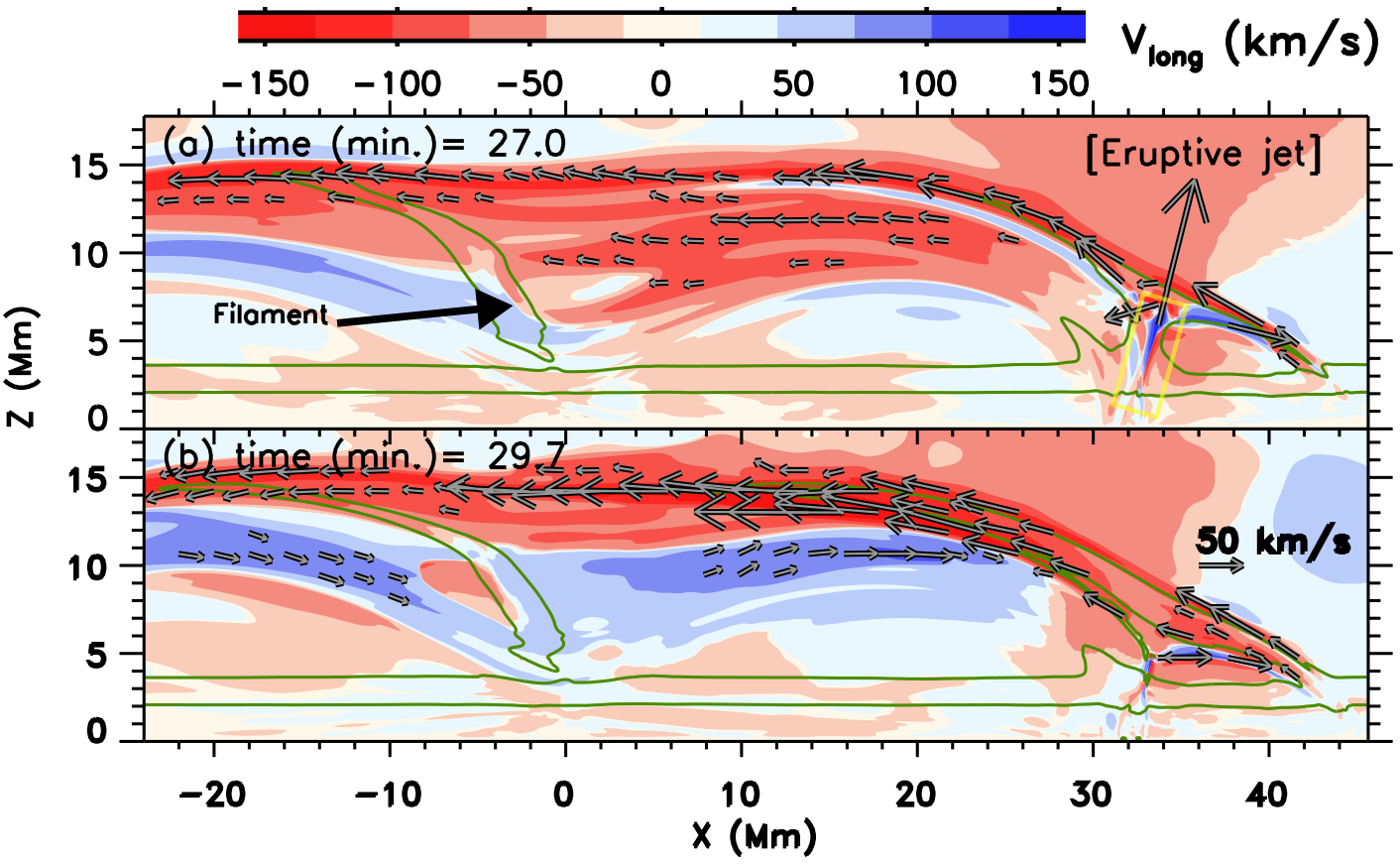}
\caption{Interaction of  the ejected plasma with a filament  channel adapted from \citet{Luna2021}. Both panels show the evolution of the jet at $t=27$ (a) and $t=29.7$(b) minutes after the start of the numerical simulation. The green contours are isolines of the density. Around x = 0 we see the green contour of the filament suspended in dips of arcades of magnetic field (indicated by a black arrow in (a)). Red and blue are the velocity flows projected along the magnetic field lines. The vector field shows the flow velocity. For a better visualisation, only vectors with a velocity greater than 20 km s$^{-1}$ are shown. To show the scale of the velocity vectors, a 50 km s$^{-1}$ arrow is shown on the right side of the panel (b).}
\label{fig:theory_flows}
\end{figure}

The nonlinear waves and jets arising in the model as a result of the
reconnection were produced in two stages; in the first one, an Alfvénic front
followed by a more standard sonic shock were launched from the BCS. These are
difficult to discern in the current observations. Then, in the second stage,
reconnection in a quasi-vertical current sheet starts below the BCS; a
twisted flux rope, resembling a mini-filament, is produced and hurled upward
against the BCS. There follows a violent eruptive jet which is launched
along the field lines of the filament channel and eventually impacts the
prominence. In the observations, we already mentioned the appearance of a
small filament-like structure which erupts below the magnetic arcade (marked
with a red arrow in Fig.~\ref{fig:zoom}) toward the BCS. More importantly,
the major eruptive jet at $\sim$20:50 UT, already discussed in Figs.~\ref{fig:zoom}b, c and d
(also in the animation, \url{AIA171_feb.mov}, and in Fig.~\ref{Feb2}) clearly
originates in one of the sides of the BCS and travels along the filament
channel toward the filament. This matches quite well the magnetic field and
velocity patterns obtained in the numerical model, as apparent in, {\it e.g.},
Fig.~12 of \cite{Luna2021}, adapted here with extra markers for easiness
of comparison (see Fig.~\ref{fig:theory_flows}). The two panels show the eruptive phase of the jet. In (a), the yellow rectangle delimits the area where the near-vertical current sheet occurs during the eruptive phase of the magnetic reconnection. The plasmoid originated in the vertical current sheet moves upwards and merges with the overlying field. The plasmoid moves with a high velocity as indicated by the almost vertical arrow in the figure. This produces a large perturbation that generates flows that propagate along the field lines that collide with the prominence (see Fig. \ref{fig:theory_flows}(b)). 
In contrast to the model, the jet in the observations shows a rotational motion along the filament channel with about
one turn along the entire length between the BCS and the filament; this leads
to a misalignment of about $30$~deg between some of the branches of the jet
and the filament spine. 
In the theoretical simulation, changes in field line
orientation were seen propagating along the FC, but no real rotation of the
plasma around the field lines of the FC could develop given the
two-dimensional (more precisely 2.5D) nature of the model. Finally, in the
numerical simulation the side of the BCS opposite to the filament also had an
outflow, albeit much weaker than the main outflow and jets. This is apparent
as a thin layer with velocity arrows pointing upward and away from the BCS
in Fig.~\ref{fig:comparison_theoretical_model_electr_current}. In the
observation, as pointed out in Section~\ref{sec:february-5}, there is also a
collimated brightening issuing from the BCS and pointing in the North
direction (marked with cyan arrows in Figs.~\ref{fig:zoom}c and d). 
This is suggestive of the presence of a secondary outflow
from the reconnection site also in the actual observed configuration. 

\section{Event 2: January 1, 2014}\label{sec:january}
\begin{figure}[ht!]
\centering
\includegraphics[width=0.49\textwidth]{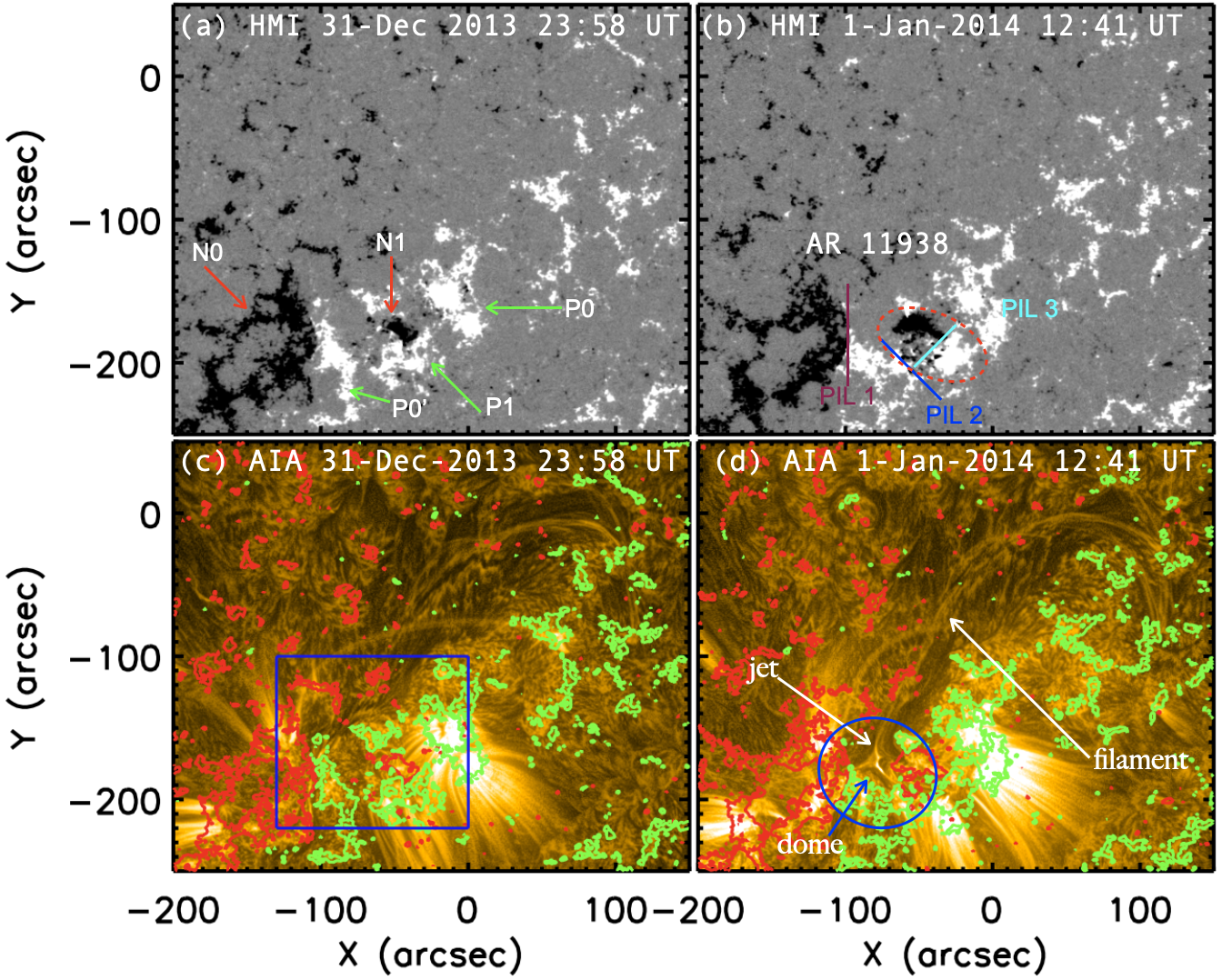}
\caption{Evolution of magnetic field (top row) and filament channel in AIA 171 \AA\ (bottom row) for event 2. The AR 11938 consists a big positive polarity patch (P0', P0, and P1) surrounded with negative polarities (N0, N1).
The different PILs (1, 2, 3) are shown in panel (b). 
The filament appears as dark areas (panel (c-d)) and the bright thin loops  are possibly arcades or magnetic field lines of the filament structure. The jet flowing towards the filament channel near the PIL2 is shown with arrow in panel (d). The dome and null-point-like structure is enclosed by a blue circle. The green and red contours in the bottom row are the HMI magnetic field contours of strength $\pm$ 50 Gauss respectively. All images are aligned together at January 1, 2014  at 15:56 UT. Blue rectangular box is the FOV for Fig. \ref{fig:Jan94}.}
\label{fig:general-view-AR-january-1}
\end{figure}

\begin{figure*}[!ht]
\centering\includegraphics[width=0.8\textwidth]{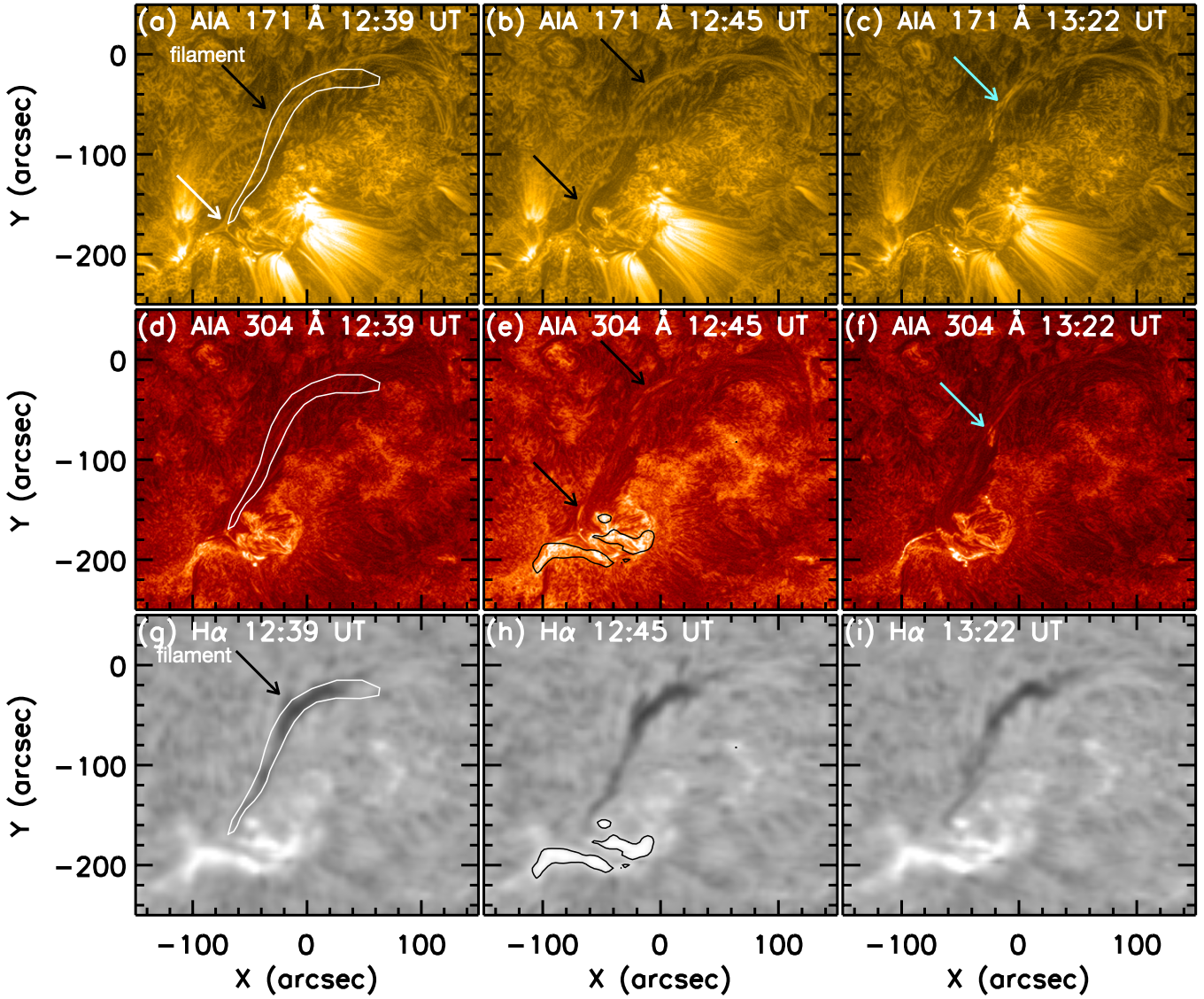}
\caption{ Filament and jet evolution on January 1, 2014 (event 2) in AIA 171 \AA\ (top row), AIA 304 \AA\ (middle row), and H$\alpha$ (bottom row). The null-point-like structure is shown with a white arrow in panel (a) at the top of the dome-like structure. The jet propagation from this null-point towards the filament channel is shown with black arrows (panel (b and e)). The cyan arrows in panel (c and f) show the later brightening associated with the jet, which reinforces the oscillations. The filament is very well observed in H$\alpha$ as a dark long structure shown with black arrow in panel indicating the sinistral handedness of the filament (g). The contours of the filament observed in H$\alpha$ are over-plotted in AIA 171 and 304 \AA\ channels (panels a,b,d). The black contours in panel e and h are for the brightening  at the jet base observed in H$\alpha$ overplotted in AIA 304 \AA. } 
\label{fig:JanEvo}
\end{figure*}

\subsection{The host active region and the filament}
On January 1, 2014, AR 11938 was located near disk center (S09W03). It
consists (Fig.~\ref{fig:general-view-AR-january-1}a, b) of disperse
polarities (P0-N0); new emerging magnetic flux (EMF; P1-N1) had
  appeared on December 31, 2013. That emergence had occurred in
the middle of the positive polarity~P0, causing it to split into two main
fragments; we call~P0' the fragment that is closer to~N0. There are several 
PILs in the active region
(Fig. \ref{fig:general-view-AR-january-1} b). PIL1 is between P0' and N0,
PIL3 between P1 and N1 in the oval of panel b. We followed the evolution of
the magnetic field for 12 hours before the eruption and observed the
expansion of P1-N1 as the flux emergence process advances; between them there
are mixed polarities probably indicating sea serpent lines which originate from the 
edge of the sunspot and create moving magnetic features \citep[MMF,][]{Dalda2008}. 
During this expansion a strong shear
occurs between~P0' and these mixed polarities. In panels c and d the coronal
emission counterpart can be seen through AIA 171 \AA images, with, superimposed,
green and red contours marking the HMI polarities. In panel d we can see a
structure resembling a dome surmounted by a null point or current sheet,
probably associated with (and resulting from) the strong
photosperic shear mentioned just above (enclosed with the blue circle).

The best identification between chromospheric, TR and coronal structures, and
the clearest view of the filament, can be gained by comparing simultaneous
H$\alpha$, AIA 304 \AA and AIA 171 \AA images, as presented in Fig.~\ref{fig:JanEvo}. 
Looking at the H$\alpha$ panels (bottom row), the
filament is clearly visible in the neighbourhood of NOAA AR 11938 and
stretches mainly in the direction SouthEast-NorthWest, with the southern tip
touching the AR, which is also where the bright jet activity described below
takes place; the top part of the filament is arched in the East-West
direction. Both from Fig.~\ref{fig:general-view-AR-january-1}c,d, and from
the AIA 171 \AA and AIA 304 \AA snapshots of Fig.~\ref{fig:JanEvo} ({\it e.g.}, panel b and
e), one sees that the filament has one end in the vicinity of the dome. The
magnetic configuration near the dome appears to be quadrupolar again, like in
the February event described in Sec.~\ref{sec:february-5}, and thus of the
type studied in the theoretical model of \citet{Luna2021}. The filament is
sinistral according to the orientation of its lateral barbs in the H$\alpha$
images (Figure \ref{fig:JanEvo} g,h,i).

\begin{figure}[!h]
\centering\includegraphics
[width=0.49\textwidth]{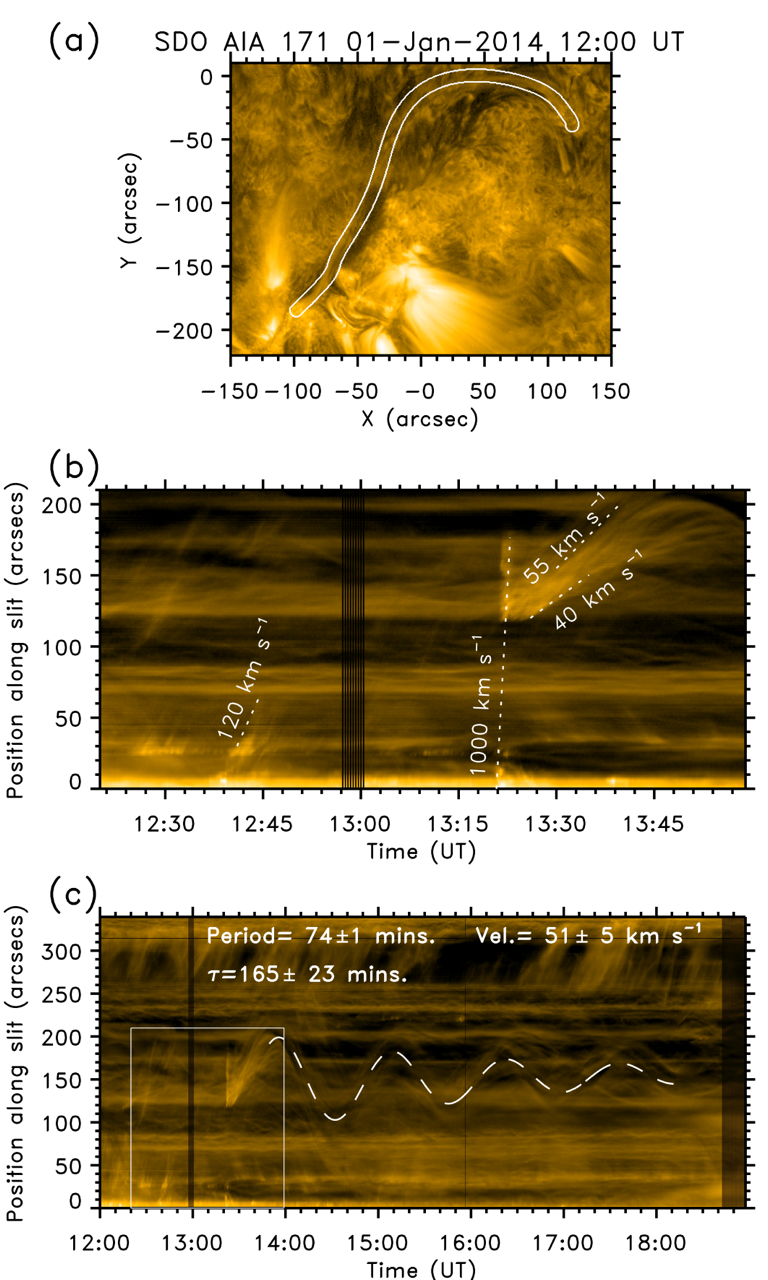}
\caption{Tracing of the oscillations in the filament for the first event on January 1, 2014 observed with AIA 171 \AA. 
Figure showing the filament seen in absorption in AIA 171 \AA. The artificial slit is shown as a white contour shown in panel (a) to construct the time-distance diagram shown in panels (b) and (c).
\label{fig:timedistances-event20140101}
}
\end{figure}

\begin{figure}[!h]
\centering\includegraphics
[width=0.45\textwidth]{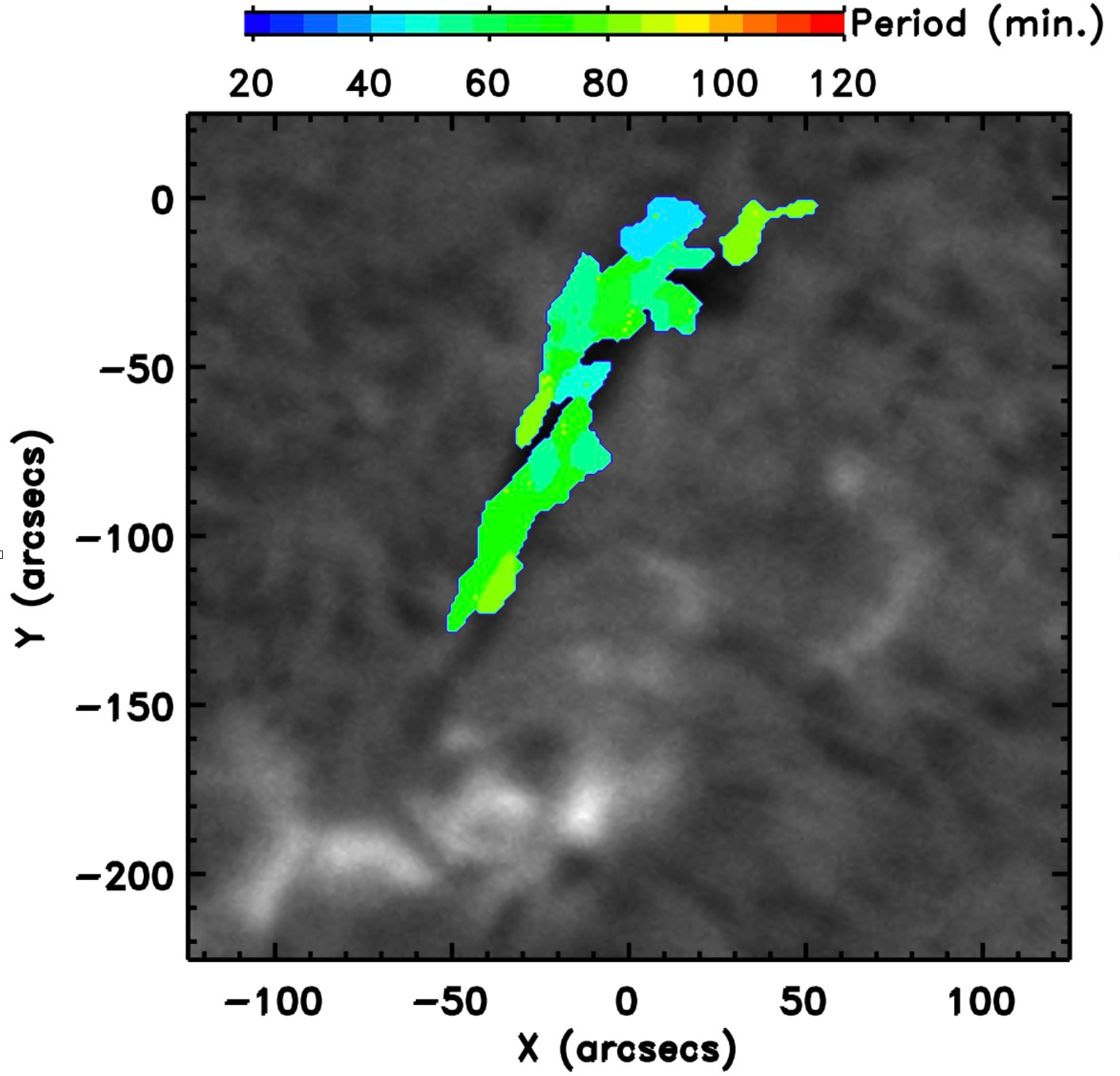}
\caption{Time periods of the filament oscillations over a GONG H$\alpha$ image on January 1, 2014 using the technique developed by \citet{luna_automatic_2022}. 
The color code shows the periods in minutes. \label{fig:periods-over-filament-secondevent}}
\end{figure}

\subsection{Jet activity}

To follow the jet propagation from its source towards the filament channel,
we analysed the AIA multi-wavelength observations in 171~and~304~\AA\ shown
in Fig.~\ref{fig:JanEvo}~(top and middle rows) and, in full for
  171~\AA, in an accompanying animation (\url{AIA171_jan.mov}). Between
12:00 to 12:35 UT, we observe many small bright jets with  short
lifetimes. They emanate from the null-point-like structure going upward in the
direction of the filament. In Fig.~\ref{fig:JanEvo} (a) we can see the hot
plasma that has emanated from these jets as bright structures on the filament
following the magnetic field lines of the filament channel.
Right after 12:40 UT a significant jet is observed: an intense brightening is seen around the area of the null point and the jet is  accompanied by apparently larger flows than the previous cases (see the left
black arrow in Figs.~\ref{fig:JanEvo} b and \ref{fig:JanEvo} e), which can also
be seen over the filament (right black arrow in Fig.~\ref{fig:JanEvo} (b)).
Prior to 13:20 UT, no clear association of the jets with filament
oscillations can be established. At 13:21 UT, however, an intense brightening (like a burst) is observed around the null point. 
This 
brightening leads to a jet propagating on the left-hand branch issuing from the null-point-like structure more than in the
  direction of the filament (cyan arrow in Figs.~\ref{fig:JanEvo} c and Fig.~\ref{fig:JanEvo} f).
Alternatively, this could be caused by activity unrelated to the null point area. For example, at 13:22 UT at x=-40  and y=25 arcsecs, a brightening occurs simultaneously with that of the null point. However, we do not observe 
any flows impacting the filament
from other regions. 

Similarly to the previous event (Sec.~\ref{sec:oscFeb} and
  \ref{subsec:evolution-jet20140205}), we can analyze the velocity of the jet
  and the oscillatory pattern using the same curved slit for both
  determinations, as shown in
  Fig.~\ref{fig:timedistances-event20140101}(a). Concentrating here on the
  jet velocity, Fig.~\ref{fig:timedistances-event20140101}(b) shows the
time-distance diagram for the AIA171 intensity along the slit in
the first $120$ minutes. The initial jet activity (say, before 12:45
  UT) is clearly seen in the form of bright features moving along the
slit. These early jets all have a similar velocity; using a bright
  feature that starts at 12:42 UT, we obtain a speed of $\sim$~120 km
s$^{-1}$, as indicated with a dotted line on the left of the panel.
Later, at 13:22~UT, we can identify an almost vertical brightening stretching between
  100$\arcsec$--150$\arcsec$ and followed by a more gradual arch-like
  collection of 
  brightenings all the way to the right end of the panel.
The inclination of the almost vertical brightening corresponds to a
propagation velocity of about $1000$~km~s$^{-1}$ [dashed line in
  Fig.~\ref{fig:timedistances-event20140101}(b)]. Continuing that dashed line
toward the origin of the slit one can see that it reaches the neighborhood of
what appears to be a counterpart of the semi vertical brightening near
the dome+null-point region. Still, it is not possible to find any causal
connection between the two brightenings: we cannot identify in any of the EUV
AIA channels anything propagating along the slit that could cause the
filament to brighten in the $100\arcsec$-$150\arcsec$ region of the slit as a
result of the activity near the null-point region. Going now to the arch-like
brightenings, we can clearly identify velocities between $40$ and
$55$~km~s$^{-1}$ in the initial part of the arches. Looking further in time
[Fig.~\ref{fig:timedistances-event20140101}(c)], one sees that the arches are
the initial part of oscillatory features in the filament (to help the
identification, the location of panel b is indicated in panel c with a white
rectangle). These are analyzed in the next section. 

\begin{figure}[!ht]
\centering\includegraphics[width=0.40\textwidth]{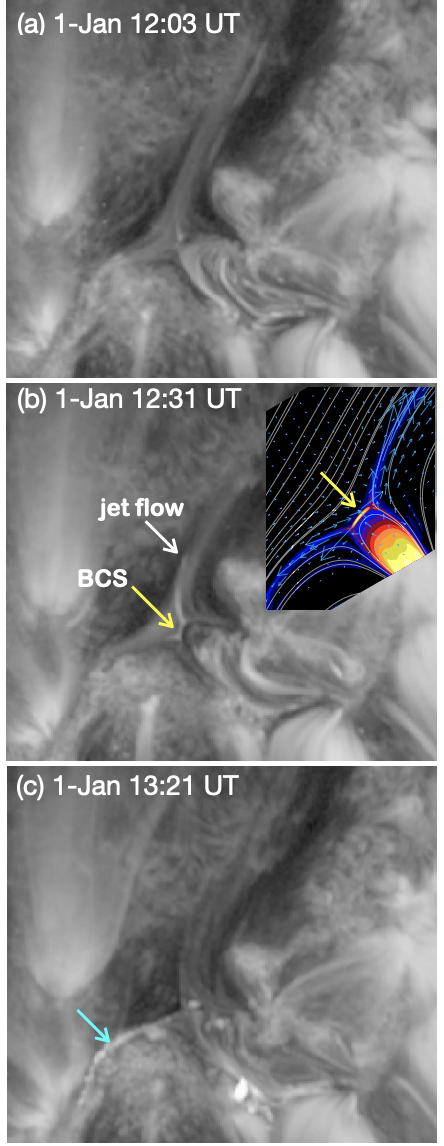}
\caption{
Zoom view of the jet base region  in 171\AA\ maps as indicated with the blue rectangular box in Fig. \ref{fig:general-view-AR-january-1} (c). The three panels show three stages of the temporal evolution of the jet. In (a) a X-point structure can be identified in the central part. This structure evolves and acquires a BCS H-type shape marked by the white arrow. Important flows are seen emanating from the H structure. The inset in panel (b) shows the electric currents distribution from the \citet{Luna2021} model (adapted from  their Fig. 4b). The white curves are the magnetic field lines and the blue vectors are the velocities of the plasma. Finally in (c) this structure disappears and a brightening is seen around it. The cyan arrow in panel (c) shows the brightening along the field lines associated with the eruptive phase. 
}
\label{fig:Jan94}
\end{figure}

\subsection{Filament oscillations}
\label{sec:oscJan}
We focus now on the oscillatory patterns that start at \hbox{$\sim$~13:20~UT} following the almost vertically propagating
brightening just discussed and last for more than 4 hours. In contrast to the February event,  the oscillations here
are more clearly discernible with all the threads oscillating more or
less in phase. The oscillation can be distinguished until 18:00~UT.
A fit to one of the oscillatory patterns gained through
Eq.~\ref{eq:sinusoid} can be seen through the white curve on the panel, with
the parameters of the fit indicated on the top right.
The period is $74\pm1$~min, which is in agreement with the results of the catalog
by \citet{luna_gong_2018}; and, the damping time is $165\pm23$ minutes,  which is almost in agreement with the catalog value of $121\pm15$ minutes when considering the error bars.
Using Eq. \eqref{eq:new-pendulum} we obtain a radius of curvature of the dipped
field lines of $R=170$ Mm. 
%
%
Also, Eq. \eqref{eq:seismology-magnetic-field} gives the minimum field value to support the cold plasma of the prominence against gravity. With the periods obtained we find that the range of minimum values of the field intensity $B_{min}=11-35$ Gauss.
As mentioned in the previous event, by studying
the oscillations with a slit we can determine the amplitude, period and
damping time. However, the slit covers a restricted area of the
filament. Using the technique introduced by \citet{luna_automatic_2022} for this
case, as we did for the February event of Sec.~\ref{sec:oscFeb}, we can see
the main oscillation periods in the whole filament (see 
Fig. \ref{fig:periods-over-filament-secondevent}).

It can be seen that almost the entire filament oscillates and that the
periods in the different regions range from 50--90 minutes but
predominantly near $80^{+10}_{-8}$~minutes. This value is in agreement with the value
obtained with the slit analysis. Using this range of periods we obtain a
range of radii of curvature $69-285$ Mm and a minimum field intensity of $7-46$~G.

\subsection{Comparison of the observations to the theoretical model}

Like for the jets discussed in
  Sec.~\ref{subsec:comparison-model-1st-event}, the Jan 1st event has
  some striking similarities to the \citet{Luna2021} model. The basic
  underlying magnetic configuration of the model described in
  Sec.~\ref{subsec:comparison-model-1st-event} is also present here, namely
  an arcade associated with a bipolar parasitic zone at one end of the FC
  surmounted by a null X-point.
Three snapshots along the time evolution that are particularly
  relevant for the comparison are shown in Fig.~\ref{fig:Jan94} using AIA
171\AA\ maps. In panel (a), taken from the map at $t=$ 12:03 UT, an
  arcade is clearly seen at the southern end of the filament: going from the
  arcade upwards, the remote connectivity of the magnetic field and its
  orientation change abruptly at a point resembling an X-crossing: all of
  this indicates the probable presence of a null point at the top of the
  arcade. 
  Photospheric shear at the feet of the
arcade causes the null point to become a BCS where reconnection takes
place. In fact, at the later time shown in panel~(b) ($t=$ 12:31 UT), at
  the location of the null point one sees a structure with a clear H-shape
  probably corresponding to the BCS. The flows emanating from it are directed
  mainly toward the filament, but also in the opposite direction. From the
  model, we expect this to correspond to the quiescent phase of the jet,
with continuous plasma flows emanating from the reconnection
  site. The resemblance to the numerical model is remarkable, as can be
  gathered by comparing the observed structure with that in the inset, adapted
  from the \citet{Luna2021} paper, which contains a color map of the module
  of the electric current with velocity arrows superimposed.
In the numerical model, the reconnection in the BCS ceases later
    on, but, by then, the arcade has become elongated in the vertical
    direction and an intense vertical current 
    sheet appears at its center; there follows reconnection across that
    current sheet, and a violent jet is ejected along the filament channel. 
%
Panel (c), taken from the AIA171 map at $t=$ 13:21 UT can correspond to this phase,
with a brightening along the field lines indicated by the cyan arrow.  That
brightening
could be associated with the eruptive phase, in which a plasmoid or mini
filament is ejected. 
The eruption as it is described in the model is not really identifiable.
 However, these brightenings indicate
that an energetic process has taken place. As an additional
  indication of the violence of the process, the oscillation of the
filament starts right after this brightening, as we have studied in
Sect. \ref{sec:oscJan}.

\section{Conclusion}
This observational study has been inspired by the scenario recently proposed
by \citet{Luna2021} for triggering the large-amplitude oscillations (LAOs) in
filaments. The authors propose that
the LAOs may be initiated by jets in a reconnection process in the
magnetic structure of the filament.
%
%
We have been able to identify two case studies which show similarities to the
model. In both cases we analyse observational data to understand how the
oscillation starts and how the filament oscillates after this
perturbation. 
In the first case, occurring on February 5, 2015 the jet perturbs a large portion of the H$\alpha$ filament that oscillates mainly with a  period of 74 min. The period detected in 171 \AA\  is not so clearly defined, in a range of periods of  $70- 110$~min.,  leading us to believe that the filament  is made by different structures at different altitudes. In the second case, on January 1, 2014, a large
  portion of the H$\alpha$ filament is perturbed as well, and the oscillations have a range
  of periods of $50-90$~min.
In both cases periods reaching 120 minutes are detected.
These values for the period are consistent with those of
typical LAOs and with the results from the catalog of
\citet{luna_gong_2018}. We  have applied 
prominence seismology using the pendulum model by
\citet{luna_effects_2012,luna_extension_2022} allowing us to compute the
curvature of the dips in the flux rope containing the filament, and the
magnetic field strength. The latter 
%
has been estimated to be of the order of 30 Gauss in both filaments. These
values are in agreement with direct measurements from observations \citep{Leroy1984,Levens2016,Levens2017}.
Both cases described in this work are intermediate filaments with one end connected to active regions and the magnetic field can be rather large \citep{Lopez2007,Carlos2019}.
\citet{Luna2017} and more recently \citet{kucera_comparison_2022} have shown that the seismological predictions of the pendulum model work well and give similar results of the field geometry and field strength as other techniques.

Both events have similarities to the \citet{Luna2021} model. Both show a quadrupolar structure with a parasitic bipole at one end of the filament channel.
%
On top of this parasitic polarity there is a classical X-shaped
structure in the AIA brightenings strongly resembling a
null-point configuration.
In both cases the parasitic polarity is produced by magnetic flux
emergence. There is also a strong shear along the polarity
inversion line which may be responsible for disturbing the null point and
producing the magnetic reconnection that leads to the jet.
A structure with the appearance of a breakout current sheet can be
identified which surprisingly resembles the structures obtained in
the \citet{Luna2021} numerical simulations.

In the first observed event (February 5, 2015), the formation
of the BCS is followed by the eruption of a mini filament. This
indicates that both the quiescent and eruptive phase of the jet
occur almost at the same time, in contrast to the model. This may be due to
the fact that the magnetic shear injection in the parasitic region is higher
than in the model. In the model we did not explore the possibility of
increasing the injection rate, which will be left for a future study.
When the jet is launched, part of it goes more or less directly to impact the
dense material in the filament, but part of it runs along the overarching
field lines. An interpretation is that the filament is sitting at the bottom
of the main dips of a flux rope, but the field lines above the filament are
arched, perhaps also moderately twisted around the filament as in
\citet{Aulanier2002}. 

In the second event (January 1, 2014), the breakout current sheet-like
structure is observed for a long time, which, as in the previous case,
is in agreement with the model. We identify this first phase as
quiescent; the 
emanating jet does not seem to impact the filament since it does not
produce oscillations. These flows probably travel along field lines
of the filament channel without threads.  As seen in the numerical modelling,
the initial height of the null point relative to the
  filament determines how the different phases of the jet impact the
filament. 
After some time, a sudden brightening around the parasitic zone is observed, which could be identified with the eruptive phase of the jet. After this brightening, the filament starts to oscillate with a large amplitude motion.
However, we cannot identify any plasma flow from the jet area to the
filament. Perhaps, the plasma moves along field lines that are
partially hidden by the filament itself or by the foreground
emission of coronal structures. 
The triggering of the oscillations in this second event could be caused by activity unrelated to the null point. However, we did not observe plasma flows reaching the filament from other regions. An alternative explanation could be that some disturbance (e.g. reconnection) produces changes in the magnetic structure of the filament that disturb its cold plasma.

Still, the direct impact of jets remains the most probable cause of the filament oscillations in the first event. Further, interestingly, a preprint published after submission of the present manuscript reports
on another event where filament oscillations are caused by jets,  also showing clear bilateral flows \citep{Tan2023}.


Nevertheless we  conclude that the events presented here bear a clear resemblance
to those described in the model by \citet{Luna2021}. A
comprehensive study of the triggering processes of 
LAOs in filaments is needed to better understand these processes. A numerical
study is also needed to investigate how the jet synthesis is affected by
different parameters such as boundary driving. It is also necessary to extend
the study to three dimensions.

\begin{acknowledgements}
We acknowledge the anonymous referee for the valuable/constructive comments and suggestions.
This research has been supported by the European Research Council through the
Synergy Grant number 810218 (``The Whole Sun'', ERC-2018-SyG) and by the
Research Council of Norway through its Centres of Excellence scheme, project
number 262622.  M.L. acknowledges support through the Ram\'on y Cajal
fellowship RYC2018-026129-I from the Spanish Ministry of Science and
Innovation, the Spanish National Research Agency (Agencia Estatal de
Investigación), the European Social Fund through Operational Program FSE 2014
of Employment, Education and Training and the Universitat de les Illes
Balears. This publication is part of the R+D+i project PID2020-112791GB-I00,
financed by MCIN/AEI/10.13039/501100011033. Funding by the Spanish Ministry
of Science, Innovation and Universities through project PG2018-095832-B-I00
is also gratefully acknowledged. 

\end{acknowledgements}

\bibliography{references-4}

\bibliographystyle{aa}
\end{document}